\documentclass[reprint,amsmath,amssymb,aps,prl,superscriptaddress]{revtex4-2}

\usepackage{graphicx}
\usepackage{dcolumn}
\usepackage{bm}
\usepackage{hyperref}
\usepackage{color}

\begin{document}

\title{Landau Damping Beyond Smooth Velocity Distributions: A Dispersion-Free Lagrangian Time-Domain Framework}

\author{Huasheng Xie}
\email{huashengxie@gmail.com}
\affiliation{Beijing VeloAlpha Technology Co., Ltd., Beijing, 100080, China}

\author{Jinsong Zhao}
\affiliation{Purple Mountain Observatory, Chinese Academy of Sciences, Nanjing 210023, China}

\date{\today}

\begin{abstract}
The classic theory of Landau damping requires the velocity distribution function (VDF) to be analytically continued into the complex plane, implicitly requiring analyticity, which imposes constraints far more severe than infinite smoothness. Yet physical plasmas---encountered in discrete simulations, noisy spacecraft measurements, or truncated fusion distributions---are inherently non-analytic. This discrepancy poses a foundational ``smoothness paradox'': why does Landau damping robustly persist in systems where the mathematical prerequisite of analyticity is profoundly violated? Here we resolve this paradox by demonstrating that wave-particle interaction is governed by a time-dependent resonance width $\Delta v_\mathrm{res}\sim 1/kt$. Using a dispersion-free Lagrangian time-domain solver, we show that this finite width kinematically coarse-grains microscopic VDF defects at early times, acting as a natural low-pass filter that validates smooth analytical proxies---explaining the observed robustness. However, as $t\to\infty$ the resonance width narrows, inevitably forcing the wave to resolve exact topological non-smoothness. This late-time resolution triggers three distinct breakdowns: VDF truncation arrests the resonant phase transition to yield undamped discrete Van Kampen modes; observational noise induces transient algebraic spikes via linear phase-space aliasing; and step-like gradients drive anomalously violent reactive instability growth. We establish the breakdown timescale $t_b\sim 1/k\delta v$, where $\delta v$ is the characteristic scale of the non-smooth defect, providing a quantitative criterion that redefines the validity boundaries of analytic continuation in kinetic theory.
\end{abstract}

\maketitle

\textit{Introduction.}---The phenomenon of collisionless wave dissipation, famously known as Landau damping~\cite{Landau1946}, stands as one of the most profound cornerstones of modern plasma physics. Mathematically, the damping rate is determined by deforming integration contours in the complex velocity plane to capture the residues of isolated poles---a procedure strictly predicated on $f_0(v)$ being an entire analytic function, such as a perfect Maxwellian.

Physical reality, however, fundamentally contradicts this idealization. High-resolution measurements from the Magnetospheric Multiscale~(MMS) and Parker Solar Probe reveal that empirical VDFs are inherently discrete and plagued by Poisson counting noise~\cite{Burch2015,Fox2016}. In magnetically confined fusion devices, auxiliary heating and runaway electron avalanches routinely create truncated, slowing-down, or step-like non-Maxwellian distributions~\cite{Heidbrink2008}. In all these non-ideal scenarios, $f_0(v)$ possesses singularities, cut-offs, or stochastic fluctuations strictly on the real axis, challenging the strict mathematical justification for analytic continuation. While theoretical frameworks exist for non-analytic distributions (e.g., yielding algebraic decay or discrete modes), and Ng~\textit{et al.}~\cite{Ng1999} demonstrated that infinitesimal collisions replace the continuous spectrum~\cite{VanKampen1955, Case1959} via velocity diffusion, providing a fully self-consistent time-domain picture linking short-term apparent damping and long-term breakdown in strictly collisionless plasmas remains elusive.

This yields the ``smoothness paradox'': if analyticity is absent in nature, why do discrete simulations and noisy spacecraft experiments still robustly exhibit macroscopic exponential damping that closely matches smooth analytical theories? 

Isolating the exact continuous-time linear dynamics of strictly non-smooth VDFs without numerical artifacts remains a formidable computational challenge. Traditional frequency-domain frameworks~\cite{Xie2013_GPDF,Xie2025_Arbitrary} target the $t\to\infty$ asymptotic roots but can suffer from ill-conditioning: projecting non-smooth VDFs onto analytic basis functions triggers Gibbs ringing. Initial-value approaches go beyond modal damping rates by resolving the evolution of wave--particle interactions. Dawson~\cite{Dawson1961}, for example, separated resonant and nonresonant electrons and derived Landau damping from the energy gained by the resonant population. While modern Eulerian Initial Value Problem (IVP) solvers~\cite{Xie2013_Residual,Filbet2001,Cheng2014} can capture intricate phase-space dynamics, mitigating numerical diffusion and preserving fine microscopic phase-space filaments ($\sim e^{-ikvt}$) over long timescales still require immense computational resources. Lagrangian Particle-in-Cell (PIC) simulations~\cite{Dawson1983} natively handle arbitrary VDFs, but purely linear responses are overwhelmed by statistical shot noise ($\sim 1/\sqrt{N}$) unless an astronomically large macro-particle count $N$ is used.

\begin{figure}[t]
    \centering
    \includegraphics[width=\columnwidth]{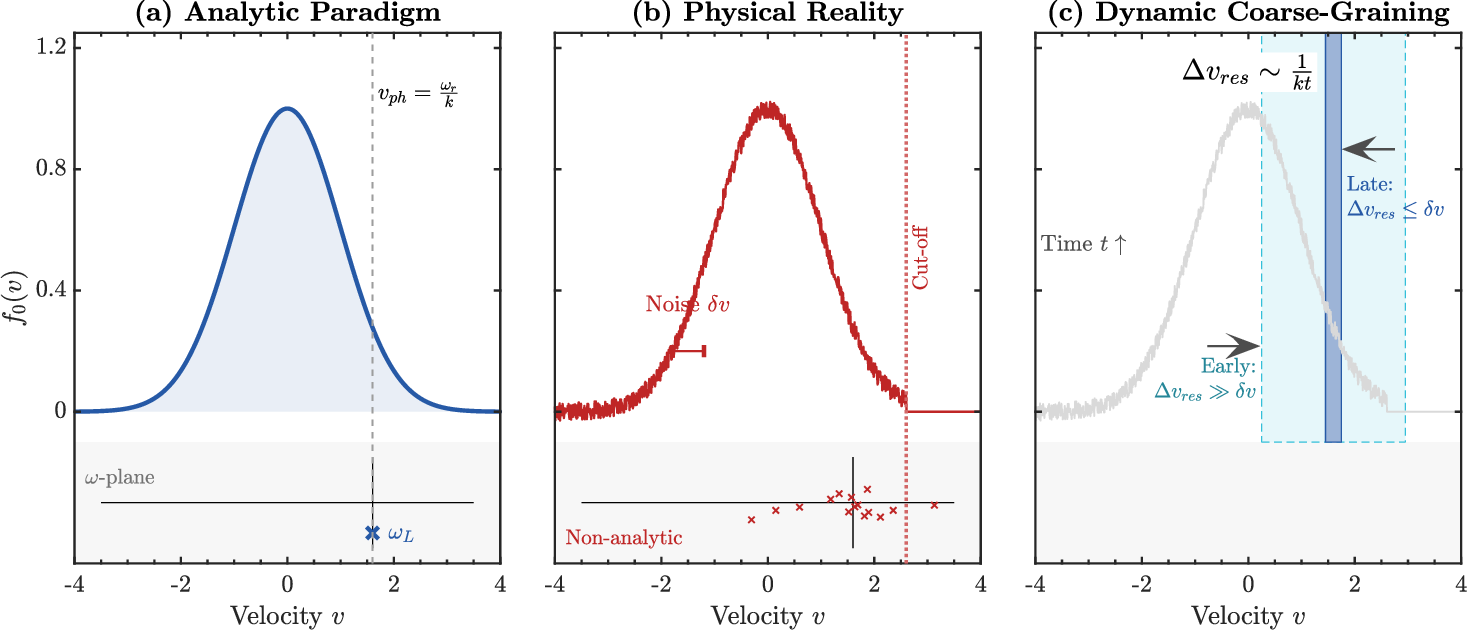}
    \caption{Conceptual resolution of the smoothness paradox.
    (a)~\emph{Analytic paradigm}: classical theory assumes infinite temporal
    resolution ($t\to\infty$), requiring an isolated pole in the lower-half
    complex plane.
    (b)~\emph{Physical reality}: discrete noise $\delta v$ and cut-offs render
    analytic continuation ill-posed; no clean pole exists.
    (c)~\emph{Dynamic coarse-graining}: wave-particle interaction operates over
    a finite window $\Delta v_\mathrm{res}\sim 1/kt$ (shaded band).
    At early times ($\Delta v_\mathrm{res}\gg\delta v$), the broad window
    averages over microscopic defects, justifying smooth proxies.
    At late times ($\Delta v_\mathrm{res}\leq\delta v$), the narrowing window
    resolves exact singularities, triggering the breakdown of damping at
    $t_b\sim 1/k\delta v$.}
    \label{fig:schematic}
\end{figure}

In this Letter, we present a dispersion-free Lagrangian time-domain framework that isolates the exact linear kinetic responses of strictly non-smooth distributions. We resolve the smoothness paradox via a dynamically narrowing resonance width (Fig.~\ref{fig:schematic}), and show that its inevitable asymptotic vanishing fundamentally shatters the Landau paradigm.

\textit{Methodology: A Dispersion-Free Lagrangian Time-Domain Framework.}---To achieve zero numerical dispersion in phase-space advection and eliminate statistical noise simultaneously, we adopt a Lagrangian displacement formulation for the 1D linear Vlasov-Poisson equations. Let $\xi(v,t)$ and $\eta(v,t)$ denote the microscopic spatial displacement and velocity perturbation of a phase-space fluid element. In Fourier space (wavenumber $k$), the linearized equations of motion are:
\begin{eqnarray}
    \partial_t \xi &=& -ikv\,\xi + \eta, \label{eq:xi} \\
    \partial_t \eta &=& -ikv\,\eta - E. \label{eq:eta}
\end{eqnarray}
The macroscopic electric field $E(t)$ is self-consistently determined by the linearized Poisson equation:
\begin{equation}
    E(t) = \int_{-\infty}^{\infty} f_0(v)\,\xi(v,t)\,dv
          \approx \sum_{j=1}^{N_v} f_0(v_j)\,\xi_j(t)\,\Delta v.
    \label{eq:poisson}
\end{equation}
The discretized VDF array in Eq.~(\ref{eq:poisson}) is physically equivalent to Dawson's multi-beam model~\cite{Dawson1960}; however, instead of diagonalizing a severely ill-conditioned $\mathcal{O}(N_v^3)$ matrix, we integrate the exact dynamics directly in the time domain at $\mathcal{O}(N_v)$ cost.

To rigorously establish that the continuous version of this Lagrangian framework captures the complete kinetic physics without omission, we prove its exact mathematical equivalence to the standard Eulerian linearized Vlasov equation:
\begin{equation}
    \partial_t\delta f + ikv\,\delta f = E\,\partial_v f_0.
    \label{eq:vlasov_euler}
\end{equation}
By Liouville's theorem and phase-space continuity, the Eulerian density perturbation $\delta f$ maps to the Lagrangian displacements via the kinematic relation:
\begin{equation}
    \delta f = -ikf_0\,\xi - \partial_v(f_0\,\eta).
    \label{eq:mapping}
\end{equation}
Substituting Eq.~(\ref{eq:mapping}) into the left-hand side of Eq.~(\ref{eq:vlasov_euler}) and expanding the convective derivative yields:
\begin{eqnarray}
    \partial_t\delta f + ikv\,\delta f
    &=& -ikf_0(\partial_t\xi + ikv\,\xi) \nonumber \\
    && -\partial_v(f_0\,\partial_t\eta) - ikv\,\partial_v(f_0\,\eta).
    \label{eq:expand}
\end{eqnarray}
We now substitute the Lagrangian governing equations~(\ref{eq:xi}) and (\ref{eq:eta}). The first parenthesis gives $\partial_t\xi + ikv\,\xi = \eta$, and $\partial_t\eta = -ikv\,\eta - E$. Equation~(\ref{eq:expand}) becomes:
\begin{eqnarray}
    \partial_t\delta f + ikv\,\delta f
    &=& -ikf_0\,\eta
        - \partial_v\!\left[f_0(-ikv\,\eta - E)\right] \nonumber \\
    &&  - ikv\,\partial_v(f_0\,\eta).
    \label{eq:almost}
\end{eqnarray}
Applying the product rule $\partial_v(ikv\,f_0\,\eta) = ikf_0\,\eta + ikv\,\partial_v(f_0\,\eta)$, the two groups of $\eta$-containing terms cancel identically, leaving exactly $\partial_v(Ef_0) = E\,\partial_v f_0$, which is Eq.~(\ref{eq:vlasov_euler}). One may further verify that integrating Eq.~(\ref{eq:mapping}) over velocity recovers the Poisson equation~(\ref{eq:poisson}), completing the proof.

This exact equivalence reveals a decisive computational advantage. Standard differential formulations often require the explicit evaluation of $\partial_v f_0$---an operation that becomes singular for step-like or truncated VDFs. While advanced weak-form solvers can mitigate this, in our Lagrangian framework, $f_0(v)$ inherently enters the dynamics \emph{only} as a continuous weighting function in the integral~(\ref{eq:poisson}); its derivative never appears explicitly. The solver is therefore intrinsically immune to non-smooth gradients. Moreover, because $f_0(v)$ is analytically separated from the microscopic displacement $\xi$, the method is entirely free of the statistical shot noise inherent in PIC simulations.

\textit{Exact Integration via ETD.}---To eliminate artificial dissipation in phase-space advection, we integrate the advection operator $-ikv$ analytically using Exponential Time Differencing~(ETD)~\cite{Cox2002}. Assuming $E$ is piecewise constant over $\Delta t$ and approximated by the Crank-Nicolson average $\bar{E}=\tfrac{1}{2}(E^n+E^{n+1})$~\cite{CrankNicolson1947}, the exact phase-space update is:
\begin{eqnarray}
    \eta^{n+1} &=& \eta^n e^{-i\omega\Delta t} - \bar{E}\,C_1(v), \label{eq:etd1}\\
    \xi^{n+1}  &=& \xi^n  e^{-i\omega\Delta t} + \eta^n\Delta t\,e^{-i\omega\Delta t} - \bar{E}\,C_2(v), \label{eq:etd2}
\end{eqnarray}
where $\omega=kv$, and the analytic coefficients are $C_1(v) = (1-e^{-i\omega\Delta t})/(i\omega)$ and $C_2(v) = (e^{-i\omega\Delta t}+i\omega\Delta t\,e^{-i\omega\Delta t}-1)/\omega^2$. Both coefficients are naturally regularized at the wave-particle resonance $\omega\to 0$ via L'H\^opital's rule: $\lim_{\omega\to 0}C_1 = \Delta t$ and $\lim_{\omega\to 0}C_2 = \Delta t^2/2$. No artificial complex frequency shift is required. Substituting Eq.~(\ref{eq:etd2}) into the discrete Poisson equation~(\ref{eq:poisson}) and solving implicitly for $E^{n+1}$ yields a simple algebraic update. ETD combined with a non-dissipative Crank-Nicolson closure ensures zero artificial numerical diffusion in phase space. Simulations presented herein typically utilize $N_v \in [250, 2000]$ velocity grid points spanning velocity domains up to $[-6, 8]\,v_\mathrm{th}$, with time steps $\Delta t \in [0.1, 0.5]\,\omega_{pe}^{-1}$.

\textit{Time-Dependent Resonance Width and Coarse-Graining.}---Classic asymptotic theory localizes kinetic responses to the single phase velocity $v_\mathrm{ph}=\omega_r/k$. In the time domain, however, the electric field at time $t$ is driven by the integral of the microscopic perturbations. The resonant response is governed by the forced Vlasov equation, which produces terms proportional to $(e^{-ikvt} - e^{-i\omega t})/(kv - \omega)$. This response function peaks strongly at $v = \omega/k$, exhibiting a primary lobe (or coherence width) of 
\begin{equation}
    \Delta v_\mathrm{res} \sim \frac{2\pi}{kt}.
    \label{eq:dvres}
\end{equation}
The wave thus effectively \emph{coarse-grains} the VDF over $\Delta v_\mathrm{res}$ at every instant, acting as a kinematic low-pass filter whose bandwidth narrows inversely with time.

\begin{figure}[t]
    \centering
    \includegraphics[width=\columnwidth]{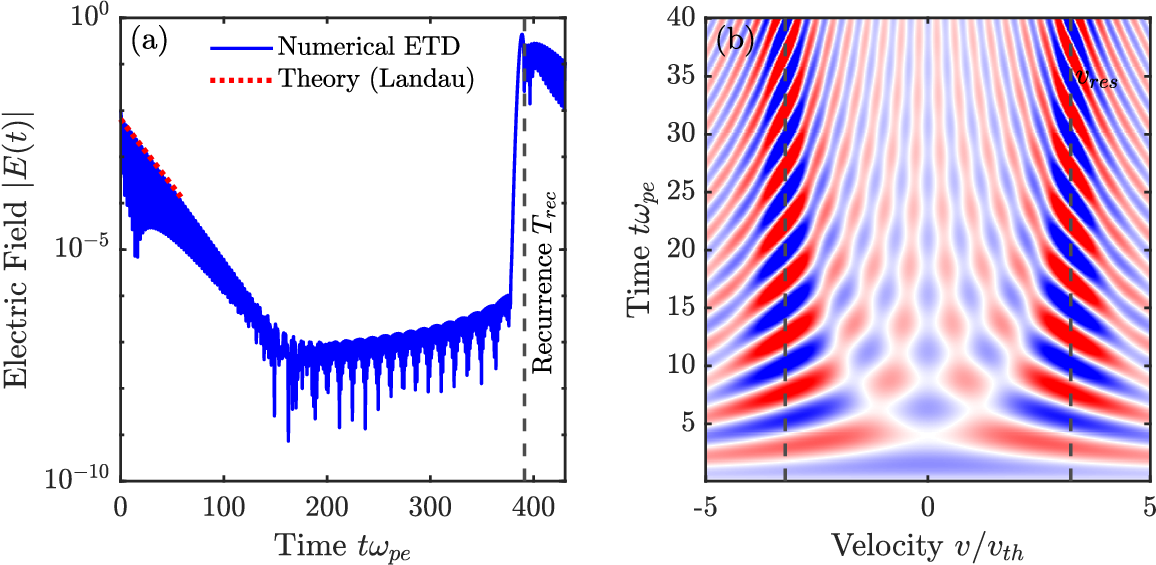}
    \caption{Smooth Maxwellian benchmark ($k=0.4$, $N_v=1500$).
    (a)~Macroscopic field envelope exhibiting exact exponential damping
    ($\gamma=-0.0661$) and a Dawson recurrence at $T_\mathrm{rec}=2\pi/k\Delta v$,
    confirming zero artificial dissipation.
    (b)~Phase-space heatmap of $\mathrm{Re}(\eta(v,t))$.
    The ever-steepening V-shaped fringes manifest kinematic phase mixing;
    the overlaid dashed lines show the theoretical resonance-width envelope
    $v=v_\mathrm{ph}\pm 1/kt$, confirming Eq.~(\ref{eq:dvres}).}
    \label{fig:fig_classic}
\end{figure}

We verify this picture with a smooth Maxwellian $f_0(v)=(2\pi)^{-1/2}e^{-v^2/2}$ at $k=0.4$ (Fig.~\ref{fig:fig_classic}). The macroscopic field exhibits exact exponential damping ($\omega_r=1.285$, $\gamma=-0.0661$) and a sharp Dawson recurrence at $T_\mathrm{rec}=2\pi/k\Delta v$~\cite{Dawson1960} (Fig.~\ref{fig:fig_classic}a), confirming the absence of artificial phase-space diffusion. The 2D phase-space heatmap (Fig.~\ref{fig:fig_classic}b) shows ever-steepening V-shaped interference fringes. The overlaid theoretical curve $\Delta v_\mathrm{res}\sim 1/kt$ precisely tracks the fringe envelope, providing a direct visual verification of Eq.~(\ref{eq:dvres}).

This establishes a quantitative failure criterion for analytic continuation. If a non-smooth VDF defect has a characteristic velocity scale $\delta v$, smooth-fitting is physically justified only while $\Delta v_\mathrm{res}\gg\delta v$. Once the wave's resolution sharpens past this scale, the approximation catastrophically fails at the breakdown time:
\begin{equation}
    t_b \sim \frac{1}{k\,\delta v}.
    \label{eq:scaling}
\end{equation}
For typical solar-wind parameters ($k\sim 0.4\,v_\mathrm{th}^{-1}$, $\delta v\sim 0.1\,v_\mathrm{th}$ set by MMS instrument energy resolution), $t_b\sim 25\,\omega_{pe}^{-1}$---well beyond most observational windows. For fusion-relevant loss-cone cut-offs ($\delta v\sim v_\mathrm{th}$), $t_b\sim 2.5\,\omega_{pe}^{-1}$, placing the breakdown squarely within experimentally accessible timescales.

\begin{figure}[t]
    \centering
    \includegraphics[width=\columnwidth]{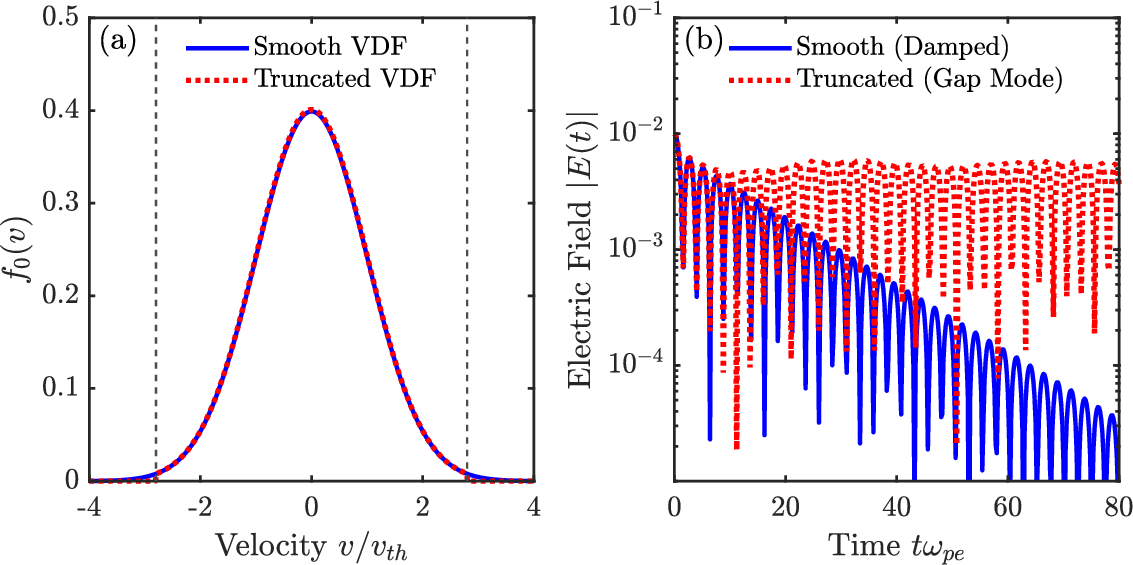}
    \caption{Topological breakdown and damping arrest.
    (a)~Smooth Maxwellian (solid) vs.\ abruptly truncated VDF at $v_c=2.8$
    (dashed).
    (b)~Macroscopic field amplitude.
    The truncated case follows the smooth proxy for $t<t_b\approx 6\,\omega_{pe}^{-1}$,
    then stabilizes into a persistent non-exponential gap mode (damping arrest).
    Smooth analytical proxies fundamentally fail to predict this
    long-term arrest of phase-mixing.}
    \label{fig:fig_trunc}
\end{figure}

\textit{Topological Defects and Arrested Damping.}---VDF truncation is a macroscopic topological defect with $\delta v\gg\Delta v_\mathrm{grid}$. We simulate a Maxwellian abruptly cut off at $v_c=2.8$, just below the resonance $v_\mathrm{ph}\approx 3.2$ (Fig.~\ref{fig:fig_trunc}a). The predicted breakdown time is $t_b\sim 1/(k\,|v_\mathrm{ph}-v_c|) \approx 1/(0.4\times 0.4)\approx 6.3\,\omega_{pe}^{-1}$.

At early times ($t<t_b$), $\Delta v_\mathrm{res}$ is broad enough to coarse-grain across the cut-off; the wave perceives a smoothed macroscopic gradient and decays in perfect agreement with the smooth proxy (Fig.~\ref{fig:fig_trunc}b). Once $\Delta v_\mathrm{res}$ narrows past $v_c$ near $t=t_b$, the exponential decay abruptly collapses and the field amplitude stabilizes into a persistent, non-exponential gap mode---a qualitative change of behavior that we refer to as \emph{damping arrest}.

The underlying mechanism reveals the ultimate failure of analytic continuation. In a smooth VDF, particles spanning the resonance velocity drive the perturbation phase through a continuous $\pi$-radian transition, enforcing coherent destructive interference and Landau damping. The truncation \emph{structurally arrests} this phase transition at $v_c$: the physical absence of particles slightly faster than the wave prevents the phase from completing the full rotation. Mathematically, the abrupt truncation implies that $\partial_v f_0$ contains a Dirac $\delta$-singularity at $v_c$. In the classical spectrum theory \cite{VanKampen1955, Case1959}, this singular gradient excites a discrete, purely oscillatory Van Kampen eigenmode with a real frequency precisely corresponding to the cut-off velocity. No analytic smoothing of the truncated data can restore the missing particles or eliminate this discrete mode, proving that macroscopic topological defects are irreducibly non-analytic.

\begin{figure}[t]
    \centering
    \includegraphics[width=\columnwidth]{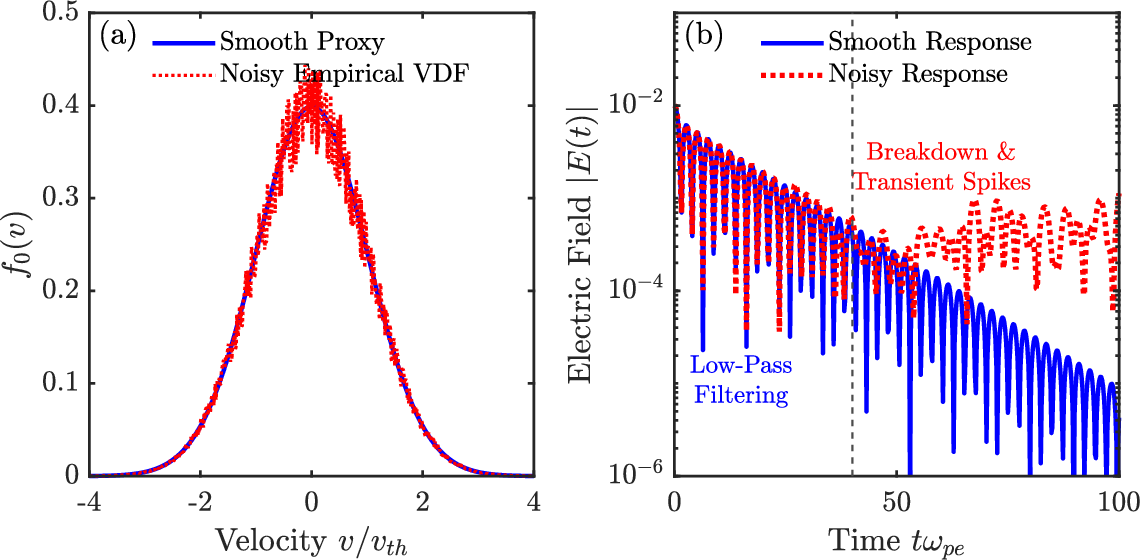}
    \caption{Kinematic low-pass filtering and noise-induced transient spikes.
    (a)~Synthetic noisy VDF (25\% Poisson fluctuations) mimicking spacecraft
    data; $\delta v = \Delta v = 12/N_v\,v_\mathrm{th}$.
    (b)~Electric field amplitude.
    Before $t_b\sim 1/k\Delta v$ (vertical dashed line), the broad resonance
    window filters the noise and the field follows the smooth proxy.
    After $t_b$, individual noise peaks are resolved, producing violent
    transient algebraic spikes via linear phase-space aliasing.}
    \label{fig:fig_noise}
\end{figure}

\textit{Observational Noise and Transient Spikes.}---Real spacecraft VDFs are plagued by Poisson counting noise. We inject 25\% random fluctuations into the Maxwellian, where $\delta v$ equals the discrete noise grid spacing $\Delta v = (v_\mathrm{max}-v_\mathrm{min})/N_v$ (Fig.~\ref{fig:fig_noise}a).

Figure~\ref{fig:fig_noise}(b) presents the dual-regime response that directly resolves the smoothness paradox. For $t<t_b$, the broad $\Delta v_\mathrm{res}$ averages out the zero-mean statistical noise; the plasma acts as a kinematic low-pass filter and the macroscopic field faithfully tracks the idealized smooth damping envelope. This mathematically justifies the engineering practice of fitting noisy empirical data to smooth functions to extract short-term wave dynamics.

For $t>t_b$, the resonance window narrows to the scale of individual noise grains, and the smoothing effect collapses. The macroscopic field erupts into severe transient spikes and erratic algebraic ringing. Because our ETD framework is strictly non-dissipative, it preserves the exact ballistic response: $E(t)$ essentially acts as a Fourier transform of the VDF evaluated at the kinematic wavenumber $k_v = kt$. Rather than artificially diffusing away high-$k_v$ structures, the wave explicitly resolves the flat, non-decaying spectrum of the white noise at large $t$, manifesting as macroscopic electric fluctuations via linear phase-space aliasing. Inferring long-term plasma stability from smoothly-fitted empirical data can therefore severely underestimate late-time transient wave activities.

\begin{figure}[t]
    \centering
    \includegraphics[width=\columnwidth]{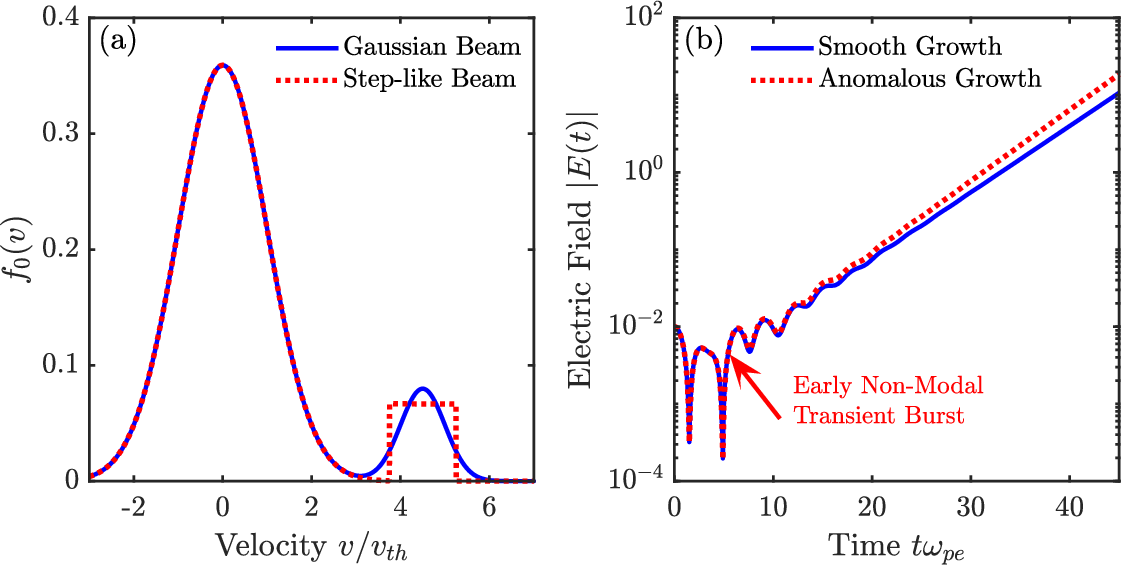}
    \caption{Two-stage instability amplification: smooth vs.\ non-smooth beam
    ($k=0.3$, $N_v=2000$, $\Delta t=0.1\,\omega_{pe}^{-1}$).
    (a)~Smooth Gaussian beam vs.\ non-smooth square beam of equivalent total
    density and characteristic width.
    (b)~Electric field growth.
    \emph{Early stage} ($t\lesssim 12\,\omega_{pe}^{-1}$): the broad resonance
    window $\Delta v_\mathrm{res}\sim 1/kt$ coarse-grains both profiles
    identically, yielding nearly coincident growth---the instability analog of
    the smoothness paradox.
    \emph{Late stage}: as $\Delta v_\mathrm{res}$ narrows onto the singular
    edge ($\partial_v f_0\to\infty$) of the square beam, the wave experiences the cold beam limit (reactive instability), driving an anomalously elevated asymptotic growth rate
    that vastly outpaces the smooth warm beam Gaussian prediction.}
    \label{fig:fig_inst}
\end{figure}

\textit{Singular Gradients and Reactive Instabilities.}---The necessity of retaining exact non-smooth dynamics extends to kinetic instabilities. We investigate a bump-on-tail instability ($k=0.3$, $N_v=2000$, $\Delta t=0.1\,\omega_{pe}^{-1}$) driven by two beams with identical total density and characteristic width but different profiles: a smooth Gaussian and a non-smooth square (step-like) beam (Fig.~\ref{fig:fig_inst}a). Both simulations start from the same small uniform displacement $\xi(v,0)$; all dynamical differences arise solely from the shape of $f_0(v)$.

Figure~\ref{fig:fig_inst}(b) reveals a striking two-stage behavior that directly mirrors the coarse-graining mechanism. \emph{At early times}, despite their vastly different profile shapes, both beams produce nearly identical macroscopic growth. This is the smoothness paradox manifested in an instability context: the broad early-time resonance window $\Delta v_\mathrm{res}\sim 1/kt$ integrates over the local VDF gradient, and because both beams carry equal total density and characteristic width, their coarse-grained averages are indistinguishable to the wave. The standard Landau growth rate $\gamma\propto\partial_v f_0|_{v_\mathrm{ph}}$ is thus the same for both at this stage.

\emph{At later times}, however, the narrowing $\Delta v_\mathrm{res}$ begins to resolve the profile shapes with increasing fidelity. The Gaussian beam has a bounded maximum slope $\max|\partial_v f_0|<\infty$, so its growth rate saturates at the smooth analytical prediction (kinetic instability). The leading edge of the square beam, by contrast, constitutes a Dirac $\delta$-singularity in $\partial_v f_0$ with infinite positive gradient. In the limit of a weak beam ($n_b \ll n_0$), this is equivalent to transitioning from a warm beam kinetic instability ($\gamma \propto n_b$) to a cold beam reactive/fluid instability ($\gamma \propto n_b^{1/3}$). As $\Delta v_\mathrm{res}$ sharpens onto this edge, the wave fully ``feels'' the cold beam limit, driving an anomalously elevated asymptotic growth rate that vastly outpaces the smooth Gaussian prediction. In the language of Eq.~(\ref{eq:scaling}), the defect scale $\delta v\to 0$ gives $t_b\to 0$, meaning the breakdown of smooth-fitting begins immediately yet only fully manifests once $\Delta v_\mathrm{res}$ has narrowed sufficiently to weight the singular edge. Replacing sharp localized injection events with smoothened analytical proxies therefore catastrophically underestimates instability onset timescales.

\textit{Discussion and Conclusion.}---We have identified the precise conditions under which analytic continuation breaks down in collisionless wave-particle interactions. By deploying a dispersion-free Lagrangian ETD solver, we established that plasma waves interact with the VDF through a dynamically narrowing resonance width $\Delta v_\mathrm{res}\sim 1/kt$, and derived the breakdown timescale $t_b\sim 1/k\delta v$ at which smooth-fitting catastrophically fails. For spacecraft-measured VDFs ($k\sim 0.4$, $\delta v\sim 0.1\,v_\mathrm{th}$), $t_b\sim 25\,\omega_{pe}^{-1}$---explaining why Landau damping appears robust in short observational windows. For fusion loss-cone distributions ($\delta v\sim v_\mathrm{th}$), $t_b\sim 2.5\,\omega_{pe}^{-1}$, squarely within experimentally accessible timescales.

The present mechanism is instructively compared with Ng~\textit{et al.}~\cite{Ng1999}, who showed that infinitesimal collisions replace the Case-Van Kampen continuum with discrete modes via velocity \emph{diffusion}. Here, by contrast, it is the \emph{sharpening} kinematic resolution of an already non-smooth VDF that halts phase-mixing---the two mechanisms are physically opposite, yet both break exponential Landau damping. Crucially, our results hold in the strictly collisionless limit, which is the more fundamental regime. Whether noise-induced spikes fully materialize in practice depends on whether $t_b \ll \tau_c$ (weakly collisional space plasmas, spikes develop) or $t_b \gg \tau_c$ (collisional edge plasmas, classical Landau is restored). Furthermore, if these non-smooth transient bursts grow rapidly, they can prematurely trigger macroscopic nonlinear particle trapping and phase-space oscillations \cite{ONeil1971} before the wave dissipates.

Our framework also has immediate practical implications for fusion energy research. Traditional codes evaluating wave heating face severe numerical pathologies when computing resonant absorption from non-smooth Fokker-Planck distribution outputs. Because the present method evaluates the exact resonant energy transfer on the real velocity axis without explicitly computing $\partial_v f_0$, it suggests a numerically stable perspective for evaluating wave-heating interactions (e.g., ICRF/ECRH power deposition) from arbitrary non-smooth distributions.

In summary: smooth-fitting remains a valid approximation for predicting short-term macroscopic wave dynamics, but retaining the exact non-smooth nature of empirical VDFs is absolutely indispensable for predicting long-term asymptotic stability.

\end{document}